\documentclass{ws-ijmpd}

\begin{document}

\markboth{Koichi Hirano}
{Observational Constraints on Phantom Crossing DGP Gravity}

%
\catchline{}{}{}{}{}
%

\title{OBSERVATIONAL CONSTRAINTS \\
ON PHANTOM CROSSING DGP GRAVITY}

\author{KOICHI HIRANO}

\address{Department of Physics, Ichinoseki National College of Technology\\
Ichinoseki 021-8511, Japan\\
hirano@ichinoseki.ac.jp}

\author{ZEN KOMIYA}

\address{Department of Physics, Tokyo University of Science\\
Tokyo 162-8601, Japan\\
zen@komiyake.com}

\maketitle

\begin{history}
\received{Day Month Year}
\revised{Day Month Year}
\comby{Managing Editor}
\end{history}

\begin{abstract}
We study the observational constraints on the Phantom Crossing DGP model. We demonstrate that the crossing of the phantom divide does not occur within the framework of the original Dvali-Gabadadze-Porrati (DGP) model or the DGP model developed by Dvali and Turner. By extending their model in the framework of an extra dimension scenario, we study a model that realizes crossing of the phantom divide. We investigate the cosmological constraints obtained from the recent observational data of Type Ia Supernovae, Cosmic Microwave Background anisotropies, and Baryon Acoustic Oscillations. The best fit values of the parameters with 1$\sigma$ (68\%) errors for the Phantom Crossing DGP model are $\Omega_{m,0}=0.27^{+0.02}_{-0.02}$, $\beta=0.54^{+0.24}_{-0.30}$. We find that the Phantom Crossing DGP model is more compatible with the observations than the original DGP model or the DGP model developed by Dvali and Turner. Our model can realize late-time acceleration of the universe, similar to that of $\Lambda$CDM model, without dark energy due to the effect of DGP gravity. In our model, crossing of the phantom divide occurs at a redshift of $z \sim 0.2$.
\end{abstract}


\section{Introduction}

General relativity is the most successful theory of gravity, which passes accurate tests in the solar system and laboratories. However, late-time accelerated expansion of the universe was indicated by measurements of distant Type Ia supernovae (SNIa)\cite{rie1998}\cdash\cite{fri2008}. It is not possible to account for this phenomenon within the framework of general relativity containing only matter and radiation. Therefore, a number of models containing ``dark energy'' have been proposed as the mechanism for the acceleration. There are currently many dark energy models including cosmological constant, scalar field, quintessence, and phantom models\cite{rat1988}\cdash\cite{kom2005}.
However, dark energy, the nature of which remains unknown, has not yet been detected. The cosmological constant is the standard candidate for dark energy. To explain the current acceleration of the universe, the cosmological constant must have an incredibly small value. But its value cannot be explained by current particle physics, and it is plagued with fine-tuning problems and the coincidence problem.
 
The mystery of the dark components is based on general relativity, and hence, it tells us that what we do not know may be the long distance behavior of gravity rather than the energy-momentum components in the universe. In this sense, cosmological observations open up a new window to study the properties of gravity on large scales.

An alternative method for explaining the current accelerated expansion of the universe is to extend general relativity to more general theories on cosmological scales. Instead of adding an exotic component, such as a cosmological constant to the right-hand side (i.e., the energy-momentum tensor) of Einstein's field equation, the left-hand side (i.e., the Einstein tensor, which is represented by pure geometry) can be modified. Typical models based on this modified gravity approach are $f(R)$ models\cite{noj2007}\cdash\cite{noj2008b} and the Dvali--Gabadadze--Porrati (DGP) model\cite{dva2000}\cdash\cite{def2002} (for reviews, see Ref.~\refcite{koy2008}).

In $f(R)$ models, the scalar curvature $R$ in the standard Einstein--Hilbert gravitational Lagrangian is replaced by a general function $f(R)$. By adopting an appropriate function phenomenologically, $f(R)$ models can account for late-time acceleration without postulating dark energy.

The DGP model is an extra dimension scenario. In this model, the universe is considered to be a brane; i.e., a four-dimensional (4D) hypersurface, embedded in a five-dimensional (5D) Minkowski bulk. On large scales, the late-time acceleration is driven by leakage of gravity from the 4D brane into 5D spacetime. Naturally, there is no need to introduce dark energy. On small scales, gravity is bound to the 4D brane and general relativity is recovered to a good approximation.

According to various recent observational data including that of SNIa\cite{ala2004}\cdash\cite{jas2006}, it is possible that the effective equation of state parameter $w_{\rm eff}$, which is the ratio of the effective pressure $p_{\rm eff}$ to the effective energy density $\rho_{\rm eff}$, evolves from being greater than $-1$ (non-phantom phase) to being less than $-1$ (phantom phase\cite{cal2002}${}^{,}$\cite{noj2003}); namely, it has currently crossed $-1$ (the phantom divide).

$f(R)$ models that realize the crossing of the phantom divide have been studied\cite{bam2009b}${}^{,}$\cite{bam2009a}. On the other hand, in the original DGP model\cite{dva2000}\cdash\cite{def2002} and a phenomenological extension of the DGP model described by the modified Friedmann equation proposed by Dvali and Turner\cite{dva2003}, the effective equation of state parameter never crosses the $w_{\rm eff}$ = $-1$ line.

In this paper, we explain the ``Phantom Crossing DGP model''\cite{hir2010} by further extending the modified Friedmann equation proposed by Dvali and Turner\cite{dva2003}. In our model, the effective equation of state parameter of DGP gravity crosses the phantom divide line. We investigate the allowed parameter region in detailed comparison with recent observational data, and show the validity and the properties of our model.

This paper is organized as follows. In the next section, we summarize the original DGP model and the modified Friedmann equation proposed by Dvali and Turner\cite{dva2003}, and demonstrate that the effective equation of state does not cross the $w_{\rm eff}$ = $-1$ line in these frameworks. Then, we review ``the Phantom Crossing DGP model'' by additionally extending the modified Friedmann equation proposed by Dvali and Turner. We show that the effective equation of state parameter of our model crosses the phantom divide line. In Section \ref{obsconst} we study the cosmological constraints on the Phantom Crossing DGP model obtained from the recent observations of SNIa\cite{kas2009}, Cosmic Microwave Background (CMB) anisotropies\cite{kom2009}, and Baryon Acoustic Oscillations (BAO)\cite{eis2005}. We show the allowed parameter region and the best fit parameters and investigate the validity and the properties of our model. Finally, the conclusions are given in Section \ref{conclu}.
\section{Model}

\subsection{DGP model}

The DGP model\cite{dva2000} assumes that we live on a 4D brane embedded in a 5D Minkowski bulk. Matter is trapped on the 4D brane and only gravity experiences the 5D bulk. 

The action is
\begin{eqnarray}
S & = & \frac{1}{16\pi}M_{(5)}^3\int_{bulk}{d^5x\sqrt{-g_{(5)}}R_{(5)}} \nonumber \\
  &   & + \frac{1}{16\pi}M_{(4)}^2\int_{brane}{d^4x\sqrt{-g_{(4)}}(R_{(4)}+L_m)},
\end{eqnarray}
where the subscripts (4) and (5) denote quantities on the brane and in the bulk, respectively. $M_{(5)}$ ($M_{(4)}$) is the 5D (4D) Planck mass, and $L_m$ represents the matter Lagrangian confined on the brane. The transition from 4D gravity to 5D gravity is governed by a crossover scale $r_c$.
\begin{equation}
r_c = \frac{M_{(4)}^2}{2M_{(5)}^3}.
\end{equation}
On scales larger than $r_c$, gravity appears in 5D. On scales smaller than $r_c$, gravity is effectively bound to the brane and 4D Newtonian dynamics is recovered to a good approximation. $r_c$ is a single parameter in this model.

Assuming spatial homogeneity and isotropy, a Friedmann-like equation is obtained on the brane\cite{def2001}${}^{,}$\cite{def2002}:
\begin{equation}
H^2 = \frac{8\pi G}{3}\rho+\epsilon\frac{H}{r_c} \label{dgp_fri},
\end{equation}
where $\rho$ is the total cosmic fluid energy density on the brane. $\epsilon = \pm 1$ represents the two branches of the DGP model. The solution with $\epsilon = +1$ is known as the self-accelerating branch. In this branch, the expansion of the universe accelerates even without dark energy, because the Hubble parameter approaches a constant, $H = 1/r_c$, at late times. On the other hand, $\epsilon = -1$ corresponds to the normal branch. This branch cannot undergo acceleration without an additional dark energy component. Hence, in what follows, we consider the self-accelerating branch ($\epsilon = +1$) only.

For the second term on the right-hand side of Eq. (\ref{dgp_fri}), which represents the effect of DGP gravity, the effective energy density is
\begin{equation}
\rho_{r_c} = \frac{3}{8\pi G}\frac{H}{r_c}, \label{rho_eff}
\end{equation}
and the effective pressure is
\begin{equation}
P_{r_c} =  -\frac{1}{8\pi G}\left(\frac{\dot{H}}{r_cH}+3\frac{H}{r_c}\right), \label{p_eff}
\end{equation}
where $\dot{H}=dH/dt$ is the differential of the Hubble parameter with respect to the cosmological time $t$.
Using Eqs. (\ref{rho_eff}) and (\ref{p_eff}), the effective equation of state parameter of DGP gravity is given by
\begin{equation}
w_{r_c} = \frac{P_{r_c}}{\rho_{r_c}}.
\end{equation}

Assuming that the total cosmic fluid energy density, $\rho$, of Eq. (\ref{dgp_fri}) contains matter and radiation, the effective equation of state of DGP, $w_{r_c}$, can also be exactly expressed as follows\cite{lue2004}${}^{,}$\cite{lue2006}:
\begin{equation}
w_{r_c} = - \frac{1}{1+\Omega_m+\Omega_r},
\end{equation}
where $\Omega_m$ and $\Omega_r$ are the normalized energy density of matter and radiation on the brane, respectively; i.e., $\Omega_m = (8\pi G/3H^2)\rho_m$ and $\Omega_r = (8\pi G/3H^2)\rho_r$. ($\rho_m\propto a^{-3}$, $\rho_r\propto a^{-4}$).

In realistic ranges of the energy density, $\Omega_m > 0$ and $\Omega_r \ge 0$, the value of the effective equation of state cannot be less than or equal to $-1$. That is, the effective equation of state never crosses the phantom divide line in the original DGP model.

\subsection{DGP model extended by Dvali and Turner}

Dvali and Turner\cite{dva2003} phenomenologically extended the Friedmann-like equation (Eq. (\ref{dgp_fri})) of the DGP model. This model interpolates between the original DGP model and the pure $\Lambda$CDM model with an additional parameter $\alpha$. The modified Friedmann-like equation is \cite{dva2003}
\begin{equation}
H^2 = \frac{8\pi G}{3}\rho+\frac{H^{\alpha}}{{r_c}^{2-\alpha}}. \label{dt_fri}
\end{equation}
For $\alpha = 1$, this agrees with the original DGP Friedmann-like equation, while $\alpha = 0$ leads to an expansion history identical to that of $\Lambda$CDM cosmology.

Differentiating both sides of Eq. (\ref{dt_fri}) with respect to the cosmological time $t$, we obtain the following differential equation:
\begin{equation}
2\dot{H}=-8\pi G(\rho+P)+\frac{\alpha\dot{H}}{(r_cH)^{2-\alpha}}, \label{hdot2}
\end{equation}
where the dot indicates the derivative with respect to the cosmological time. The quantity $P$ is the total cosmic fluid pressure on the brane.

For the second term on the right-hand side of Eq. (\ref{dt_fri}), which represents the effect of DGP gravity, the effective energy density is
\begin{equation}
\rho_{\alpha}=\frac{3}{8\pi G}\frac{H^{\alpha}}{{r_c}^{2-\alpha}}, \label{rho_dt}
\end{equation}
and from Eq. (\ref{hdot2}), the effective pressure is
\begin{equation}
P_{\alpha}=-\frac{1}{8\pi G}\left[\frac{\alpha\dot{H}}{(r_cH)^{2-\alpha}}+3\frac{H^{\alpha}}{{r_c}^{2-\alpha}}\right]. \label{p_dt}
\end{equation}
From Eqs. (\ref{rho_dt}) and (\ref{p_dt}), the effective equation of state parameter of the DGP model extended by Dvali and Turner is given by
\begin{equation}
w_{\alpha} = \frac{P_{\alpha}}{\rho_{\alpha}}. \label{w_dt}
\end{equation}

\begin{figure}[h!]
\centerline{\psfig{file=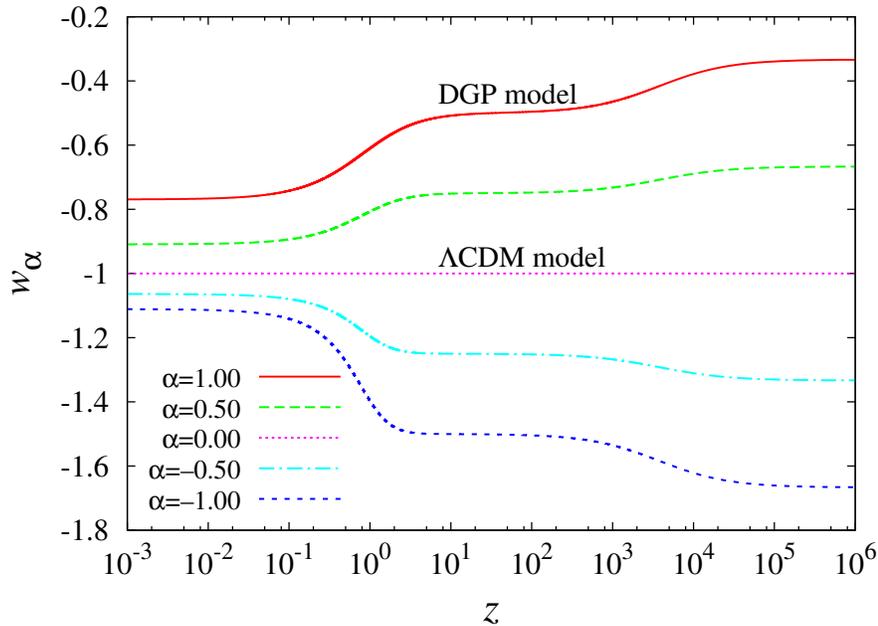,width=12.2cm}}
\caption{Effective equation of state, $w_{\alpha}$, versus redshift $z$ of the DGP model extended by Dvali and Turner for $\alpha = 1.00, 0.50, 0.00, -0.50$, and $-1.00$ (top to bottom) assuming $\Omega_{m,0} = 0.30$. For $\alpha = 1.00$, this agrees with the original DGP model, while $\alpha = 0.00$ corresponds to the $\Lambda$CDM model. \label{fig:dt_wa}}
\end{figure}

Fig. \ref{fig:dt_wa} shows a plot of the behavior of the effective equation of state of the DGP model proposed by Dvali and Turner, $w_\alpha$, versus the redshift $z$ for $\alpha = 1.00, 0.50, 0.00, -0.50$, and $-1.00$, assuming $\Omega_{m,0} = 0.30$, (The subscripts $0$ designate the present value.)

During the earlier radiation-dominated epoch ($z\gg10^4$), the effective equation of state (Eq. (\ref{w_dt})) can also be represented by the following equation with $\alpha$ \cite{dva2003}:
\begin{equation}
w_\alpha = -1 + \frac{2\alpha}{3},
\end{equation}
and during the matter-dominated era ($10^2$\hspace{0.3em}\raisebox{0.4ex}{$>$}\hspace{-0.75em}\raisebox{-.7ex}{$\sim$}\hspace{0.3em}$z$$\gg$$1$),
\begin{equation}
w_\alpha = -1 + \frac{\alpha}{2}.
\end{equation}
At present, $w_\alpha$ is a stationary value close to $-1$.

However, when $\alpha$ is positive, the effective equation of state, $w_\alpha$, will exceed $-1$ at all times. For negative $\alpha$, $w_\alpha$ is always less than $-1$. For $\alpha = 0$, $w_\alpha$ is $-1$ at all times. Based on this analysis, crossing of the phantom divide does not occur in the DGP model extended by Dvali and Turner.

\subsection{Phantom Crossing DGP model}

We propose the ``Phantom Crossing DGP model''\cite{hir2010}${}^{,}$\cite{hir2010b}, which additionally extends the modified Friedmann equation (Eq. (\ref{dt_fri})) proposed by Dvali and Turner. Our model can realize crossing of the phantom divide line for the effective equation of state of the DGP gravity.

As mentioned in the previous section, the effective equation of state parameter of the DGP model proposed by Dvali and Turner, $w_\alpha$, takes the value of greater than $-1$ for positive $\alpha$ and less than $-1$ for negative $\alpha$. When $\alpha = 0$, $w_\alpha$ becomes $-1$.
On the basis of these results, by changing the sign of $\alpha$ (from positive to negative), the effective equation of state parameter varies from being greater than $-1$ to being less than $-1$.

Here, we shall assume that the physics that modifies Friedmann-like equation satisfies the following simple requirement:
\begin{itemize}
\item The sign of $\alpha$ in Eq. (\ref{dt_fri}) varies from being positive to being negative {\bf keeping the model as simple as possible}. 
\end{itemize}
In accordance with this requirement, we make the following assumption\cite{hir2010}:
\begin{equation}
\alpha = \beta - a, \label{beta}
\end{equation}
where $a$ is the scale factor (normalized so that the present day value is unity). The quantity $\beta$ is a constant parameter. In the period when the scale factor $a$ is less than the parameter $\beta$ ($\alpha > 0$), the effective equation of state exceeds $-1$. At the point when the scale factor $a$ equals $\beta$, ($\alpha = 0$), the equation of state's value will be $-1$. In the period when the scale factor $a$ exceeds the parameter $\beta$ ($\alpha < 0$), the equation of state will be less than $-1$. In this way, crossing of the phantom divide is realized in our model.

Replacing $\alpha$ by $\beta - a$ in Eq. (\ref{dt_fri}), the Friedmann-like equation in our model is given by
\begin{equation}
H^2 = \frac{8\pi G}{3}\rho+\frac{H^{\beta-a}}{{r_c}^{2-(\beta-a)}}. \label{hirano_fri}
\end{equation}
Differentiating both sides of Eq. (\ref{hirano_fri}) with respect to the cosmological time $t$, the following differential equation is obtained:
\begin{equation}
2\dot{H}=-8\pi G(\rho+P)+\frac{(\beta-a)\dot{H}-\dot{a}H\ln{(r_cH)}}{(r_cH)^{2-(\beta-a)}}. \label{hirano_hdot2}
\end{equation}
For the second term on the right-hand side of Eq. (\ref{hirano_fri}), representing the effect of DGP gravity, the effective energy density is
\begin{equation}
\rho_{\beta}=\frac{3}{8\pi G}\frac{H^{\beta - a}}{{r_c}^{2-(\beta - a)}}, \label{rho_hirano}
\end{equation}
and from Eq. (\ref{hirano_hdot2}), the effective pressure is
\begin{equation}
\hspace*{-7mm} P_{\beta}=-\frac{1}{8\pi G}\left[\frac{(\beta-a)\dot{H}-\dot{a}H\ln{(r_cH)}}{(r_cH)^{2-(\beta-a)}}+3\frac{H^{\beta-a}}{{r_c}^{2-(\beta-a)}}\right]. \label{p_hirano}
\end{equation}
Using Eqs. (\ref{rho_hirano}) and (\ref{p_hirano}), the effective equation of state of our model is given by
\begin{equation}
w_{\beta} = \frac{P_{\beta}}{\rho_{\beta}}. \label{w_hirano}
\end{equation}

We integrate the differential equation (Eq. (\ref{hirano_fri}) or Eq. (\ref{hirano_hdot2})) in the direction from present to past. Hence, the "initial" conditions are given by the parameters designating the present value, $\Omega_{m,0}$, $\beta$, and $H_0$. The value of $r_c$ is determined so that Eq. (\ref{hirano_fri}) is satisfied at the present day.

\begin{figure}[h!]
\centerline{\psfig{file=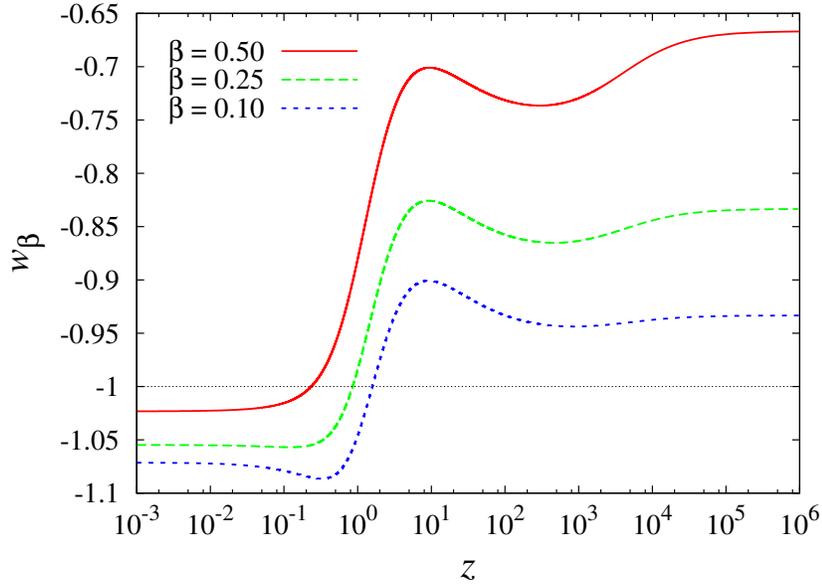,width=11.4cm}}
\caption{Effective equation of state of our model, $w_{\beta}$, versus redshift $z$. Red (solid), green (dashed), blue (dotted) lines represent the cases of $\beta = 0.50, 0.25$, and $0.10$, respectively (assuming $\Omega_{m,0} = 0.30$). \label{fig:dgp_hirano}}
\end{figure}

\begin{figure}[h!]
\centerline{\psfig{file=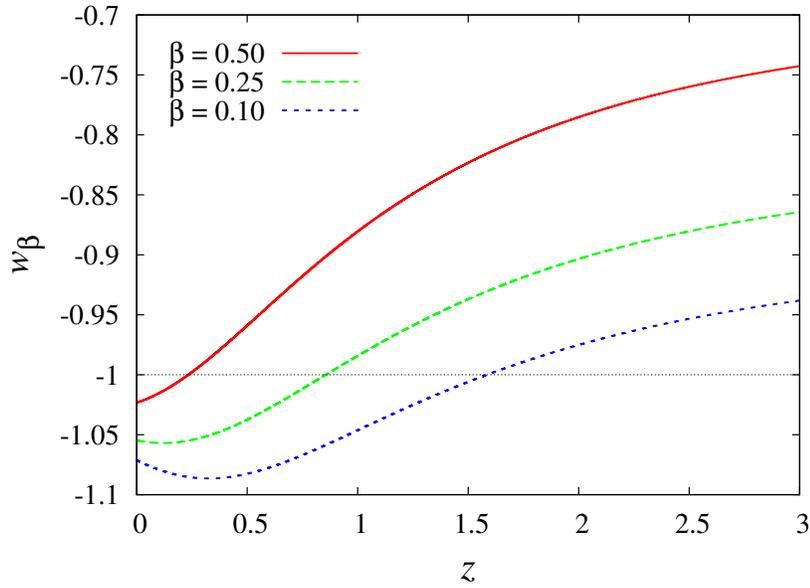,width=11.4cm}}
\caption{Detail of the behavior of $w_{\beta}$ depicted in Fig. \ref{fig:dgp_hirano} near recent epochs. \label{fig:dgp_hirano_kakudai}}
\end{figure}

Fig. \ref{fig:dgp_hirano} shows a plot of the effective equation of state of our model, $w_\beta$, versus the redshift $z$ (also refer Fig. \ref{fig:dgp_hirano_kakudai}, which shows an enlarged view of this diagram). Our model is an extension of the DGP model and realizes crossing of the phantom divide. The effective equation of state, $w_\beta$, of models for $\beta = 0.50, 0.25,$ and $0.10$ (assuming $\Omega_{m,0} = 0.30$) crosses the phantom divide line when the redshift $z \sim 0.2, 0.8,$ and $1.6$, respectively. We find that the smaller the parameter $\beta$ is, the older the epoch is in which the crossing of the phantom divide occurs. $\beta$ is not necessarily equal to the scale factor at the time of crossing the phantom divide, even though Eq. (\ref{beta}) is assumed.

The recent observational data for SNIa\cite{rie2007} show that crossing of the phantom divide line occurs at a redshift $z \sim 0.2$\cite{ala2004}\cdash\cite{wuy2006}.
In our model, for $\beta = 0.50$ (when $\Omega_{m,0} = 0.30$), crossing of the phantom divide occurs at $z \sim 0.2$.

However, the redshift at the time of phantom crossing is model-dependent. In the next section, to test the validity of the Phantom Crossing DGP model, we will compare the model and recent observational data in detail.

\section{Observational Constraints \label{obsconst}}

In this section, we study the cosmological constraints on the Phantom Crossing DGP model obtained from SNIa\cite{kas2009}, CMB anisotropies\cite{kom2009}, and BAO\cite{eis2005} observations. We carry out detailed investigation of the allowed parameter region, and show the validity and the properties of our model. In what follows, we consider the spatially flat Universe (k = 0) only.
\subsection{Observational data} 
\subsubsection{Type Ia supernovae}
For the SNIa data, we use the nearby+SDSS+ESSENCE+SNLS+HST set of 288 SNIa by Kessler et al. (2009)\cite{kas2009}. We adopt the set that used SALT-II\cite{guy2007} for SNIa lightcurve-fitting, as the MLCS method\cite{phi1993}\cdash\cite{jha2007} appears to introduce some systematic biases\cite{kas2009}. The dataset gives the distance modulus at redshift $\mu_{obs}(z_i)$.

The theoretical (for a given model) distance modulus is defined by
\begin{equation}
\mu(z)=5\log_{10}{D_L}+\mu_0,
\end{equation}
where $D_L$ is the Hubble free luminosity distance given by
\begin{equation}
D_L=(1+z)\int^z_0\frac{H_0}{H(z^{\prime})}dz^{\prime},
\end{equation}
and $\mu_0$ is
\begin{equation}
\mu_0=5\log_{10}{\left(\frac{{H_0}^{-1}}{Mpc}\right)} + 25 = 42.38 - 5\log_{10}{h}, \label{eq:mu}
\end{equation}
$h$ being the Hubble constant $H_0$ in units of $100~{\rm km~s^{-1}~Mpc^{-1}}$.

We will minimize the statistical $\chi^2$ function (which determines likelihood function of the parameters) of the model parameters. For the SNIa data,
\begin{equation}
\chi_{\rm SNIa}^2=\sum^{288}_{i=1}\left[\frac{\mu(z_i) - \mu_{obs}(z_i)}{\sigma_{\mu}(z_i)}\right]^2,
\end{equation}
where $\sigma_{\mu}$ is the distance-modulus uncertainty including an additional intrinsic dispersion of $\sigma^{\rm int}_{\mu}$ = 0.14 mag\cite{kas2009}.

Because the nuisance parameter $\mu_0$ (Eq. (\ref{eq:mu})) is model-independent, we analytically marginalize it as follows:
\begin{equation}
\chi_{\rm SNIa}^2=a-\frac{b^2}{c},
\end{equation}
where
\begin{equation}
a=\sum^{288}_{i=1}\frac{[\mu(z_i) - \mu_{obs}(z_i)]^2}{\sigma_{\mu}^2(z_i)},
\end{equation}
\begin{equation}
b=\sum^{288}_{i=1}\frac{\mu(z_i) - \mu_{obs}(z_i)}{\sigma_{\mu}^2(z_i)},
\end{equation}
and
\begin{equation}
c=\sum^{288}_{i=1}\frac{1}{\sigma_{\mu}^2(z_i)}.
\end{equation}
We use the $\chi_{\rm SNIa}^2$ function in combination with that of CMB and BAO data.

\subsubsection{Cosmic microwave background}

The CMB shift parameter is one of the least model-dependent parameters extracted
from the CMB data. Because this parameter involves the large redshift behavior ($z\sim 1000$), it gives a complementary bound to the SNIa data ($z \hspace{0.3em}\raisebox{0.4ex}{$<$}\hspace{-0.75em}\raisebox{-.7ex}{$\sim$}\hspace{0.3em} 2$). The shift parameter $R$ is defined as
\begin{equation}
R=\sqrt{\Omega_{m,0}}\int_{0}^{z_{CMB}}{\frac{H_0}{H(z)}dz},
\end{equation}
where $z_{CMB}$ is the redshift at recombination. Although $z_{CMB}$ depends on the matter density $\Omega_{m,0}$ and on the baryon density $\Omega_{b,0}$ at the $\sim$ 2 percent level, we fix this redshift to the 5-year WMAP maximum-likelihood value of $z_{CMB}$ = 1090\cite{kom2009}.

Using the 5-year WMAP data of $R_{obs} = 1.710 \pm 0.019$\cite{kom2009}, the $\chi^2$ function for CMB is
\begin{equation}
\chi_{\rm CMB}^2=\left[\frac{R - 1.710}{0.019}\right]^2.
\end{equation} 

\subsubsection{Baryon acoustic oscillation}

Observations of large-scale galaxy clustering provide the signatures of the BAO. We use the measurement of the BAO peak in the distribution of luminous red galaxies (LRGs) observed in the Sloan Digital Sky Survey (SDSS)\cite{eis2005}. It gives
\begin{equation}
A_{obs}=0.469\times\left(\frac{0.96}{0.98}\right)^{-0.35}\pm 0.017,
\end{equation}
and the parameter A is calculated by
\begin{equation}
A=\sqrt{\Omega_{m,0}}\left(\frac{H_0}{H(z_{BAO})}\right)^{1/3}\left[\frac{1}{z_{BAO}}\int_0^{z_{BAO}}\frac{H_0}{H(z)}dz\right]^{2/3}
\end{equation}
and $z_{BAO} = 0.35$. The $\chi^2$ function for BAO is
\begin{equation}
\chi_{\rm BAO}^2=\left[\frac{A - 0.469\times\{(0.96/0.98)^{-0.35}\}}{0.017}\right]^2.
\end{equation}

To determine the best value and the allowed region of parameters, we will use the maximum likelihood method and need to minimize the following quantity:
\begin{equation}
\chi^2=\chi_{\rm SNIa}^2+\chi_{\rm CMB}^2+\chi_{\rm BAO}^2. \label{chi2}
\end{equation}

\subsection{Numerical results}

We present our main results of constraints from the observational data described in the previous subsection.

\begin{figure}[th!]
\centerline{\psfig{file=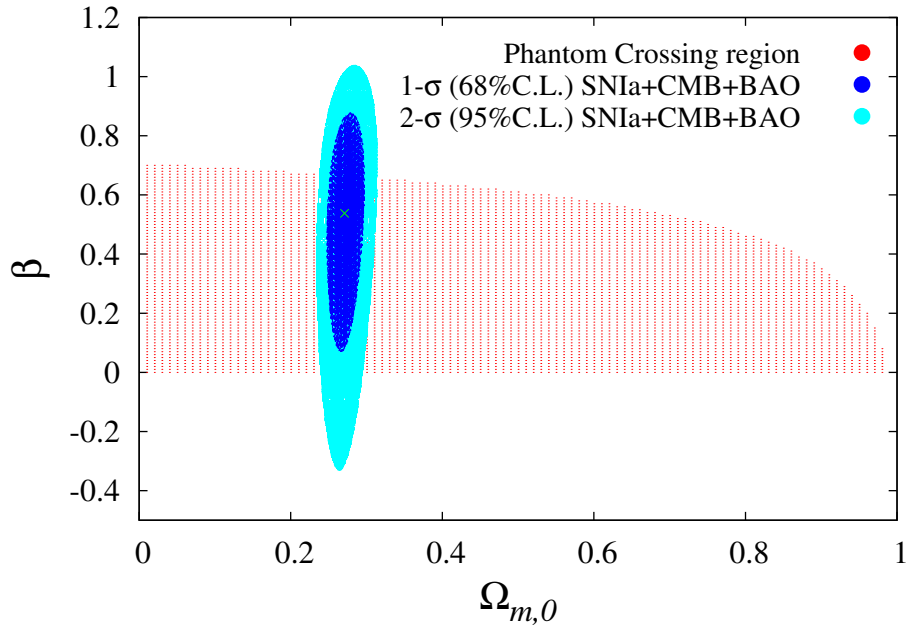}}
\caption{Probability contours in the ($\Omega_{m,0}$, $\beta$)-plane for the Phantom Crossing DGP model. Crossing of the phantom divide occurs in the red region. The blue and light blue contours show the $1\sigma$ (68\%) and $2\sigma$ (95\%) confidence limits, respectively, from a combined analysis of the SNIa, CMB, and BAO data. The point denotes the best fit. \label{d2con}}
\end{figure}

In Fig. \ref{d2con}, we plot the probability contours in the ($\Omega_{m,0}$, $\beta$)-plane for the Phantom Crossing DGP model. In the red region, crossing of the phantom divide occurs. In other words, the effective equation of state of the DGP gravity varies from being greater than $-1$ to being less than $-1$. The blue and light blue contours show the $1\sigma$ (68\%) and $2\sigma$ (95\%) confidence limits, respectively, from a combined analysis of the SNIa, CMB, and BAO data. We find that much of the $1\sigma$ (68\%) confidence level (C.L.) region is included in the region where the crossing of the phantom divide occurs.

\begin{figure}[h!]
\centerline{\psfig{file=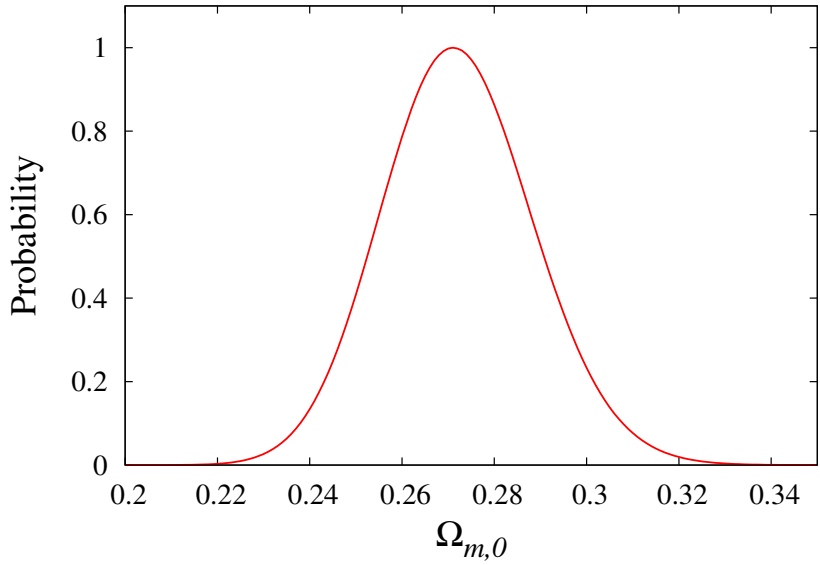,width=11.6cm}}
\caption{One dimensional probability distribution of parameter $\Omega_{m,0}$ for the Phantom Crossing DGP model from the combination of SNIa, CMB, and BAO data. \label{d1omega}}
\end{figure}

\begin{figure}[h!]
\centerline{\psfig{file=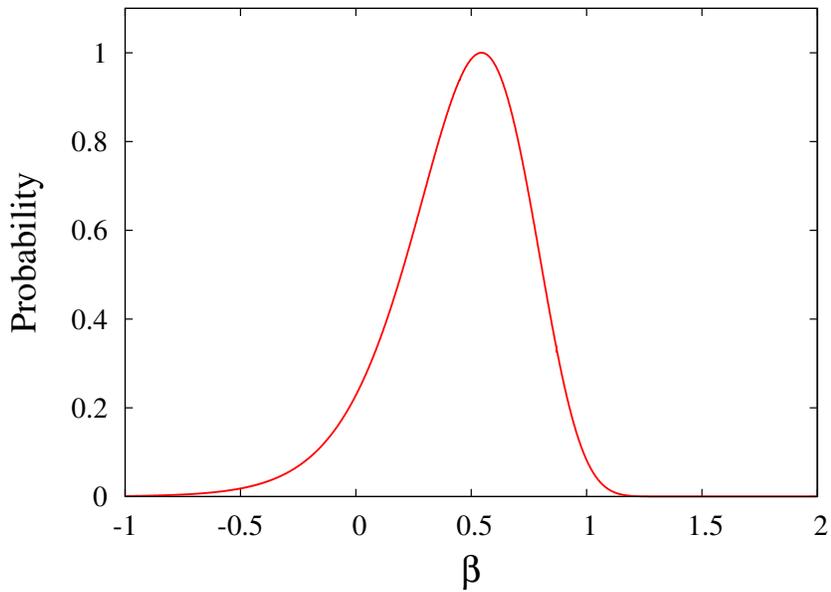,width=11.6cm}}
\caption{1D probability distribution of parameter $\beta$ for the Phantom Crossing DGP model from the combination of SNIa, CMB, and BAO data. \label{d1beta}}
\end{figure}

Fig. \ref{d1omega} shows the probability distribution of the energy density parameter of matter $\Omega_{m,0}$ for the Phantom Crossing DGP model from the combination of SNIa, CMB, and BAO data, where parameter $\beta$ is marginalized. Fig. \ref{d1beta} shows the probability distribution of parameter $\beta$ for the Phantom Crossing DGP model from the combination of SNIa, CMB, and BAO data, where parameter $\Omega_{m,0}$ is marginalized. 

The best fit values of the parameters with 1$\sigma$ (68\%) errors are
\begin{equation} 
\Omega_{m,0}=0.27^{+0.02}_{-0.02}
\end{equation}
\begin{equation}
\beta=0.54^{+0.24}_{-0.30}  \label{beta68}
\end{equation}

The best fit value of $\Omega_{m,0}$ for the Phantom Crossing DGP model is similar to the value indicated in the $\Lambda$CDM model. We obtained the comparatively stringent constraint for parameter $\Omega_{m,0}$. In Eq. (\ref{beta}), the sign of $\beta - a$ varies from being positive to being negative by the present day for $0 < \beta < 1$, because the scale factor $a$ is normalized so that the present day value is unity. We find that the 68\% C.L. region of $\beta$ (Eq. (\ref{beta68})) is included in the range of $0 < \beta < 1$.

In Table \ref{table1}, we list the best fit parameters, the $\chi^2$ values (Eq. (\ref{chi2})), and the differences of the Akaike information criterion (AIC)\cite{aka1974} and the Bayesian information criterion (BIC)\cite{sch1978}, for the $\Lambda$CDM model, the Phantom Crossing DGP model, the DGP model proposed by Dvali and Turner, and the original DGP model, from a combined analysis of the SNIa, CMB, and BAO data. The definitions of AIC and BIC are
\begin{equation}
{\rm AIC} = -2\ln{L} + 2k,
\end{equation}
\begin{equation}
{\rm BIC} = -2\ln{L} + k\ln{N},
\end{equation}
where $L$ is the maximum likelihood, $k$ is the number of free model parameters, and $N$ is the number of data points used in the fit. The $\chi^2$ value for the Phantom Crossing DGP model is the smallest in the four models. We find that the Phantom Crossing DGP model is more compatible with the observations than the original DGP model or the DGP model developed by Dvali and Turner. However, the values of AIC and BIC for the $\Lambda$CDM model are smaller than those for the Phantom Crossing DGP model, because the number of parameters of the $\Lambda$CDM model is one less than that of the Phantom Crossing DGP model.

\begin{table}[h!]
\tbl{Results of observational tests from the combination of SNIa, CMB, and BAO data.}
{\begin{tabular}{l l l l l}
\hline\hline
Model  &  Best fit parameters  &  $\chi^2$  &  $\Delta$AIC  &  $\Delta$BIC \\
\hline
$\Lambda$CDM  & $\Omega_{m,0}=0.27$  &  247.262  &  0.000  &  0.000 \\ 
Phantom Crossing DGP\cite{hir2010}${}^{,}$\cite{hir2010b} &  $\Omega_{m,0}=0.27$,~ $\beta=0.54$  &  246.729  &  1.467  &  5.137 \\
DGP by Dvali and Turner\cite{dva2003}~~~  &  $\Omega_{m,0}=0.27$,~ $\alpha=0.12$~  &  247.086~  &  1.824~  &  5.494 \\
Original DGP\cite{dva2000} &  $\Omega_{m,0}=0.29$  &  269.002  &  21.740  &  21.740 \\
\hline\hline
\end{tabular} \label{table1}}
\end{table}

The observational constraints on the DGP model developed by Dvali and Turner have been studied in Ref.~\refcite{xia2009}. Although observational data used in our paper are not the same as those they used, the result that the best fit value of parameter $\alpha$ is closer to 0 than 1 is consistent with that of Ref.~\refcite{xia2009}.

\begin{figure}[h!]
\centerline{\psfig{file=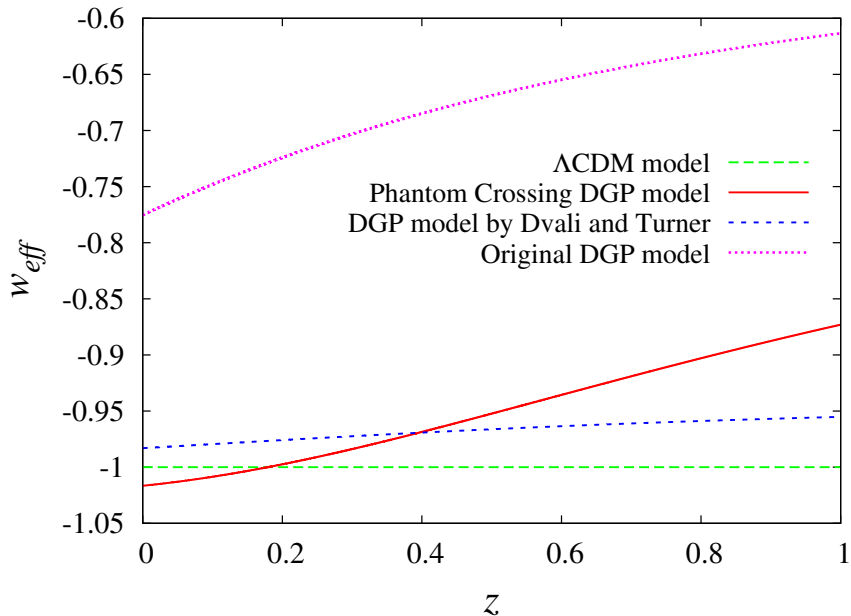,width=12cm}}
\caption{The effective equation of state of DGP gravity (or the cosmological constant), versus the redshift $z$, for the best fit models listed in Table \ref{table1}: (1) $\Lambda$CDM model, $\Omega_{m,0}$ = 0.27; (2) Phantom Crossing DGP model, $\Omega_{m,0}$ = 0.27, $\beta$ = 0.54; (3) DGP model proposed by Dvali and Turner, $\Omega_{m,0}$ = 0.27, $\alpha$ = 0.12; (4) Original DGP model, $\Omega_{m,0}$ = 0.29. \label{fig:weff}}
\end{figure}

Fig. \ref{fig:weff} shows the effective equation of state of DGP gravity (or the cosmological constant) versus the redshift $z$, for the best fit models listed in Table \ref{table1}. Only our Phantom Crossing DGP model can realize crossing of the phantom divide line at the redshift $z \sim 0.2$.

%
%

\begin{figure}[h!]
\centerline{\psfig{file=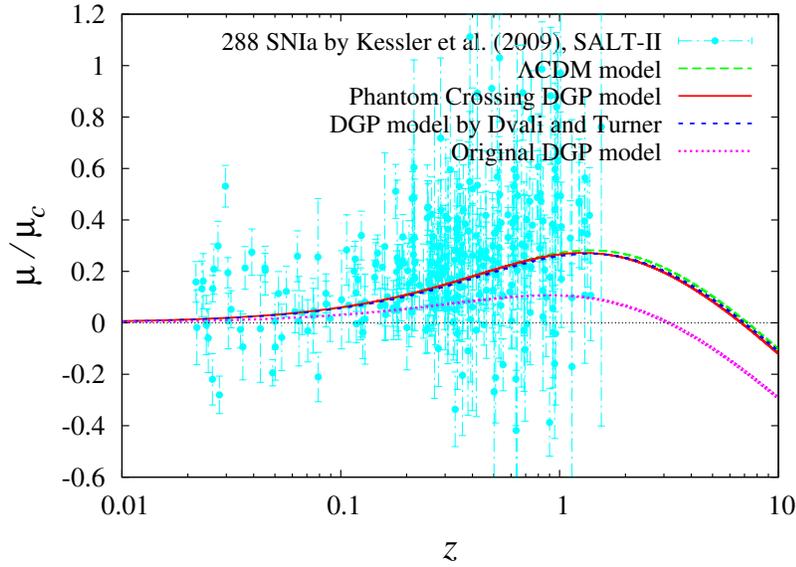,width=11.0cm}}
\caption{The distance modulus $\mu$ relative to that of a constant expansion cosmology $\mu_{c}$, versus the redshift $z$ for the best fit models listed in Table \ref{table1} (the models same as Fig. \ref{fig:weff}), and the combined sample of 288 SNIa
. \label{fig:snia}}
\end{figure}

Fig. \ref{fig:snia} shows the distance modulus $\mu$ relative to that of a constant expansion cosmology $\mu_{c}$, versus the redshift $z$ for the best fit models listed in Table \ref{table1}, and the combined sample of 288 SNIa\cite{kas2009} with the SALT-II lightcurve fitter. When $\mu /\mu_{c}$ is positive, cosmic expansion is accelerating. The Phantom Crossing DGP model can realize late-time acceleration of the universe similar to that for the $\Lambda$CDM model, without dark energy (and also for the DGP model proposed by Dvali and Turner). We will distinguish these by calculating the perturbation in future work.

\begin{figure}[h!]
\centerline{\psfig{file=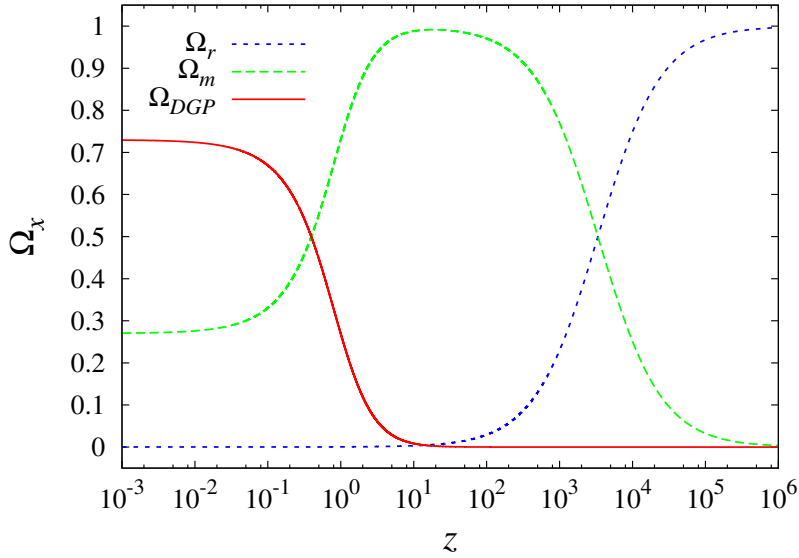,width=11.0cm}}
\caption{The normalized energy density of radiation $\Omega_r$, matter $\Omega_m$, and DGP gravity $\Omega_{DGP}$, versus the redshift $z$ in the Phantom Crossing DGP model with the best fit parameter in Table \ref{table1} ($\Omega_{m,0}$ = 0.27, $\beta$ = 0.54). \label{fig:omega}}
\end{figure}

Fig. \ref{fig:omega} shows the normalized energy density of radiation $\Omega_r$, matter $\Omega_m$, and DGP gravity $\Omega_{DGP}$ versus the redshift $z$ in the Phantom Crossing DGP model with the best fit parameters in Table \ref{table1}, where $\Omega_{DGP} = (8\pi G/3H^2)\rho_{\beta}$. $\rho_{\beta}$ is the effective energy density of DGP gravity defined by Eq. (\ref{rho_hirano}). We find that the universe is DGP gravity-dominated near recent epochs. Therefore, in the Phantom Crossing DGP model, the late-time acceleration is driven by the effect of DGP gravity not by dark energy.

\section{Conclusions \label{conclu}}

We demonstrated that crossing of the phantom divide cannot occur in the DGP model developed by Dvali and Turner or the original DGP model. We studied the Phantom Crossing DGP model by extending the DGP gravity in the framework of an extra dimension scenario. Our model realizes crossing of the phantom divide. We investigated the cosmological constraints obtained from recent observations. The best fit values of the parameters with 1$\sigma$ (68\%)
errors for the Phantom Crossing DGP model are $\Omega_{m,0}=0.27^{+0.02}_{-0.02}$,
$\beta=0.54^{+0.24}_{-0.30}$.
We find that the Phantom Crossing DGP model is more compatible with the recent observations than the original DGP model or the DGP model developed by Dvali and Turner. Our model can realize late-time acceleration of the universe very similar to that of the $\Lambda$CDM model, without dark energy due to the effect of DGP gravity. In our best fit model ($\Omega_{m,0}$ = 0.27, $\beta$ = 0.54), crossing of the phantom divide occurs at the redshift $z \sim 0.2$. It is known that the self-accelerating branch of solutions in the original DGP model suffer from ghost-like instabilities\cite{koy2007}${}^{,}$\cite{koy2005}. We will study whether the Phantom Crossing DGP model has the ghost problem in future work.



\section*{Acknowledgements}
The authors would like to thank the anonymous reviewer for their helpful comments and discussions.




\end{document}